\documentclass[showpacs,prb,aps,twocolumn]{revtex4-1}

\pdfoutput=1

\usepackage{amsmath,amsfonts,amssymb,bm,graphicx,hyperref,color,units}

\newcommand{\Pa}{\mathcal{P}}
\newcommand{\T}{\mathcal{T}}
\newcommand{\Sym}{\mathcal{S}}

\newcommand{\llangle}{\langle\!\langle}
\newcommand{\rrangle}{\rangle\!\rangle}
\newcommand{\bra}[1]{\langle #1|}
\newcommand{\ket}[1]{|#1\rangle}

\begin{document}

\title{Cumulants of time-integrated observables of closed quantum systems and $\Pa\T$-symmetry, with an application to the quantum Ising chain}

\author{James M. Hickey}
\author{Emanuele Levi}
\author{Juan P. Garrahan}

\affiliation{School of Physics and Astronomy, University of Nottingham,
Nottingham, NG7 2RD, United Kingdom}

\date{\today}

\begin{abstract}
We study the connection between the cumulants of a time-integrated observable of a quantum system and the $\Pa\T$-symmetry properties of the non-Hermitian deformation of the Hamiltonian from which the generating function of these cumulants is obtained.  This non-Hermitian Hamiltonian can display regimes of broken and of unbroken $\Pa\T$-symmetry, depending on the parameters of the problem and on the counting field that sets the strength of the non-Hermitian perturbation.  This in turn determines the analytic structure of the long-time cumulant generating function (CGF) for the time-integrated observable.  We consider in particular the case of the time-integrated (longitudinal) magnetisation in the one-dimensional Ising model in a transverse field.  We show that its long-time CGF is singular on a curve in the magnetic field/counting field plane that delimits a regime where $\Pa\T$-symmetry is spontaneously broken (which includes the static ferromagnetic phase), from one where it is preserved (which includes the static paramagnetic phase).  In the paramagnetic phase, 
conservation of $\Pa\T$-symmetry implies that all cumulants are sub-linear in time, a behaviour usually associated to the absence of decorrelation.
\end{abstract}

\maketitle

%------------Introduction------------------------%
\section{Introduction}
\label{sec:Intro}

In recent work we considered the properties of cumulants of time-integrated quantities in closed quantum systems, focusing in particular on the case of the time-integrated transverse magnetisation in the quantum Ising chain  \cite{Hickey2013}.  The aim was to highlight the importance of time-integrated operators, as opposed to time-local ones, as appropriate observables for the characterisation of real time dynamics of many-body quantum systems.  The reason is that very often many-body systems, both classical and quantum, display collective dynamical behaviour of a more complex nature that what is suggested by their structural or stationary properties.  Glass forming systems are a typical example of this \cite{Chandler2010,*Biroli2013}. 
For such systems time-integrated observables, in particular those which are extensive in both system size and time, 
serve as dynamical order parameters, as their higher moments can encode the full range of dynamical fluctuations. 
This is essentially the idea behind full-counting statistics (FCS) both in mesoscopics \cite{Levitov1993,Levitov1996,Nazarov2003,Nazarov2003b,Pilgram2003,Flindt2008,Flindt2009} and in quantum optics \cite{Gardiner2004,Esposito2009,Garrahan2010}, and of the thermodynamics of trajectories formalism in stochastic systems \cite{Eckmann1985,Ruelle2004,Merolle2005,Lecomte2007,Hedges2009}.  

The main point of \cite{Hickey2013} was to show that by considering time-integrated observables it is possible to uncover the existence of distinct ``dynamical phases'' delimited by singularities of their cumulant generating function (CGF).  These transitions are the closed system equivalent of the ``trajectory transitions'' of open classical and quantum systems \cite{Garrahan2007,Garrahan2010}, i.e., singularities in the CGF of time-integrated dynamical observables indicative of far-from-Gaussian behaviour of their corresponding probability distributions (analogous to the singularities of free-energies and the non-Gaussian behaviour of order parameter distributions at equilibrium phase transitions).   For time-integrated observables of closed quantum systems there is in general no associated probability distribution, but
the CGF does exist, and it can be non-analytic in the long time limit as a function of its argument (the ``counting'' field) \cite{Hickey2013}.  These singularities are in general extensions of static quantum phase transition points, and share many similarities with them; for example if one quenches  across these phase boundaries the echo dynamics can display ``dynamical phase transitions'' \cite{Hickey2014} similar to the ones found for quenches across static critical points \cite{Heyl2013}.

The CGF of a time-integrated observable is obtained from the spectrum of a non-Hermitian perturbation of the Hamiltonian of the system \cite{Hickey2013}.  In particular, the long-time CGF is related to the complex eigenvalue of this non-Hermitian Hamiltonian with the largest imaginary part.  Singularities in the CGF then emerge due to the closing of a gap in this complex energy spectrum.
Recently there has been a growing interest around a special class of
operators which, while non-Hermitian, do have a real spectrum (and thus could be the actual Hamiltonians of physical quantum systems)\cite{Bender1998,Bender2002,Bender2004,Bender2007,Mostafazadeh2010}.
That such operators can display a real spectrum is due to an associated $\Pa\T$-symmetry~\cite{Bender2007}.  When this $\Pa\T$ symmetry is broken these non-Hermitian operators in general have complex eigenvalues \cite{Bender1998,Bender2007}, but when the symmetry is unbroken the spectrum is real.

In this paper we explore the connection between the cumulants of time-integrated observables of a closed quantum system and the $\Pa\T$-symmetry properties of the non-Hermitian Hamiltonian from which their CGF is derived.  In certain cases, and for certain observables, the non-Hermitian Hamiltonian can display regimes of broken and of unbroken $\Pa\T$-symmetry, and these regimes then correspond to distinct dynamical phases of the system.  We apply this analysis to study the behaviour of the time-integrated longitudinal magnetisation in the one-dimensional transverse field Ising model (TFIM).  We show that the static disordered phase of the TFIM belongs to a $\Pa\T$-symmetric regime, so that the CGF of the time-integrated magnetisation is vanishing (implying that all its cumulants are sub-linear in time).  In contrast the static ordered phase belongs to a regime where $\Pa\T$ is spontaneously broken, and where the cumulants of the time-integrated magnetisation grow superlineraly with time. 

The paper is organized as follows.  In Sec.~\ref{sec:Theo} we provide the necessary theoretical background, both summarizing the method of Ref.\ \cite{Hickey2013} for studying time-integrated observables, and briefly reviewing
known results on $\Pa\T$-symmetries in spin models.  
In Sec.~\ref{sec:single} we present a warm up example to illustrate our approach, that of the time-integrated magnetisation of a single spin system and its connection to $\Pa\T$-symmetry breaking.  Our main results on the behaviour of the cumulants of the time-integrated longitudinal magnetisation in the TFIM are described in 
Sec.~\ref{sec:Ising}.  
Finally, in Sec.~\ref{sec:Conc} we give our conclusions.

%------------Background Theory-------------------%
\section{Theoretical Background}
\label{sec:Theo}

\begin{figure*}
\includegraphics[width=1.8\columnwidth]{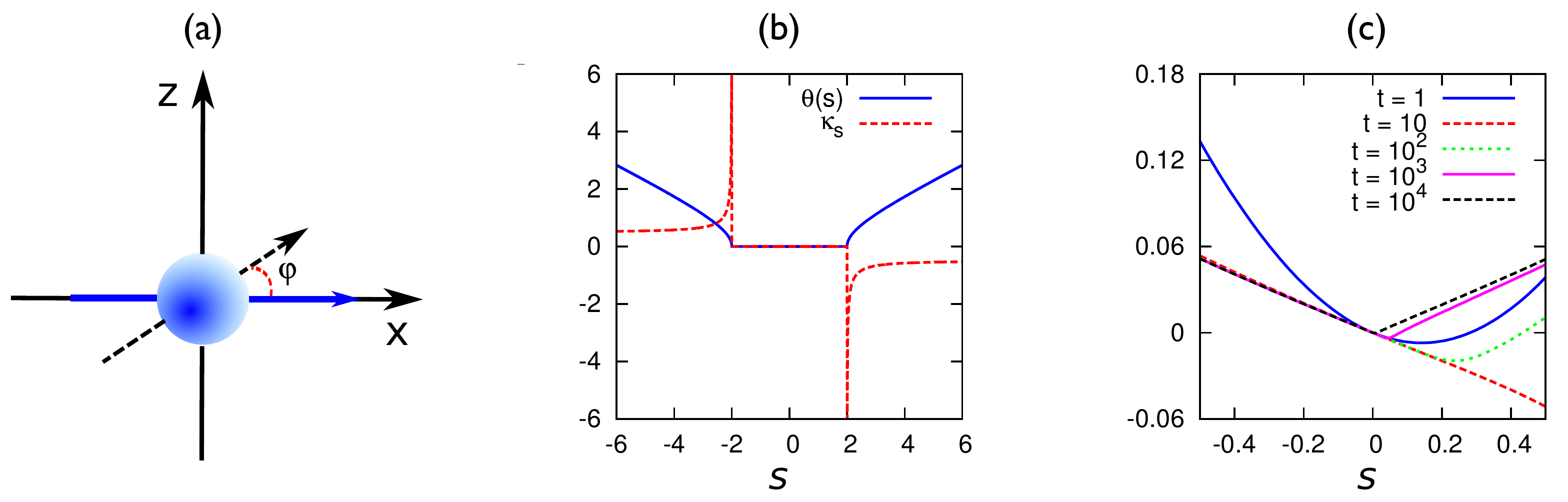}
\caption{(Color Online) (a)  Single spin which precesses around the
$x$-direction.  The time-integrated magnetisation of interest is at
an angle $\varphi$ to the precession axis. (b) Time-integrated $z$-magnetisation: for the case $\varphi=\pi/2$, the CGF $\theta(s)$ is zero for $|s| \le 2\epsilon$ and non-zero elsewhere.  The singularities of $\theta(s)$ at $\pm 2 \epsilon$ correspond to the breaking of the $\Pa\T$ symmetry of $H_{s}$.  These are also points at which $\kappa_s$ is discontinuous. (c)  The emergence of the
singularity of $\theta(s)$ at $s = 0$ in the regime where $\Pa\T$ is not present, $\varphi \neq \pi/2$ (here  $\varphi = 0.1$).}
\label{fig:fig1}
\end{figure*}

\subsection{Generating Functions of Time-Integrated Observables}
\label{sec:Form}

We consider a closed quantum system which evolves unitarily under a Hamiltonian $H$ (note we set $\hbar = 1$),  and we focus on a general
time-integrated dynamical observable whose moments (and cumulants) we are interested in:
\begin{equation}
K_{t} \equiv \int_0^{t} k_{t'}~ dt' , 
\label{Kt}
\end{equation}
where $k$ is some static system observable, and $k_{t} = U_{t}^{\dagger} k U_{t}$ is written in the Heisenberg picture with $U_{t} \equiv {e}^{-itH}$.  
The cumulants of $K_{t}$ encode information on the dynamical fluctuations of the system.  In order to obtain the cumulants, we first need to define the moment generating function (MGF) and its associated non-unitary evolution operator $T_{t}(s)$ \cite{Hickey2013},
\begin{equation}
\label{eq:deform}
T_{t}(s) \equiv {e}^{-itH_{s}}, \;\;\; H_{s} \equiv H- \frac{is}{2}k,
\end{equation}
where we deformed $H$ to a non-Hermitian Hamiltonian $H_{s}$.
With this modified evolution operator one can show that the MGF is given by \cite{Hickey2013}
\begin{equation}
\label{eq:MGF}
Z_{t}(s) = \langle T^{\dag}_{t}(s) T_{t}(s) \rangle .
\end{equation}
From these definitions it is easy to demonstrate that the moments of $K_{t}$ are obtained via differentiation, 
$\langle K_{t}^{n} \rangle = (-)^{n} \partial_{s}^{n} Z_{t}(s) |_{s \to 0}$. The CGF is given by the logarithm of 
the MGF, $\Theta_{t}(s) \equiv \log Z_{t}(s)$, such that 
$\llangle K_{t}^{n} \rrangle = (-)^{n} \partial_{s}^{n} \Theta_{t}(s)  |_{s \to 0}$, where $\llangle \cdot \rrangle$ indicates cumulant. 
These quantities define the FCS~\cite{Levitov1993,Levitov1996,Nazarov2003,Nazarov2003b,Pilgram2003,Flindt2008,Flindt2009} of the observable $K_{t}$ in this system, where 
in contrast to the usual approach used in studying FCS we consider the parameter $s$ to be real.  

To study the analytic properties of this generating function in the long-time limit it is useful to consider the scaled form of the CGF,
 \begin{equation}
 \label{eq:CGF}
 \theta(s) = \lim_{t\rightarrow \infty} \frac{\Theta_{t}(s)}{t}.
 \end{equation}
For extensive systems, such as the TFIM studied below, we also want to scale the function in order to define the large size limit, such that 
$\tilde{\theta}(s)\equiv\lim_{N\rightarrow\infty}N^{-1}\theta(s)$.   It is convenient also to define the ``order parameter'' $\kappa_{s} \equiv -\theta'(s)$, i.e.,  the long-time average of the observable $K_{t}$, per unit time, that one would obtain by controlling $s$ (rather than the actual dynamical average which corresponds to when $s=0$), and the associated susceptibility $\chi_{s} \equiv\theta''(s)$, which help to characterise the dynamical phases delimited by singularities of $\theta(s)$ \cite{Hickey2013,Hickey2014}.

\subsection{$\Pa\T$-symmetry and non-Hermitian Hamiltonians}
\label{sec:symmetry}
In standard formulations of quantum mechanics there is a strict requirement of Hermiticity of the Hamiltonian ($H^{\dagger} = H$) which ensures the 
existence of a real energy spectrum.  
Recently, after the seminal work of Bender et al.~\cite{Bender1998}, the has been much interest in the study of non-Hermitian Hamiltonians which may still have a real spectrum provided they are invariant under a space-time reflection~\cite{Bender2002,Weigert2003,Bender2004,Fring2009,Mostafazadeh2010}.
Representing the parity operator by  $\Pa$, and time reversal by $\T$, this space-time symmetry is called $\Pa\T$-symmetry, and a $\Pa\T$ symmetric Hamiltonian is defined as
\begin{equation}
\label{eq:symm}
H = H^{\Pa\T}.
\end{equation}
The operator $\Pa\T$ satisfies $(\Pa\T)^{2} = 1$,  and $\T$ is an antilinear operator~\cite{Wigner1960}.
These properties, together with condition \eqref{eq:symm} allow for a formulation of a physical theory of quantum mechanics without violating any of the original axioms.  
The condition  \eqref{eq:symm} ensures that the $\Pa\T$ operator commutes with the Hamiltonian.
A conventional $\Pa\T$ operator acts as a combination of a spatial reflection (say, $\textbf{x} \rightarrow -\textbf{x}$) and a time reversal, usually corresponding to complex conjugation.
This operator is not linear, hence the eigenstates of $H$ may differ from those of $\Pa\T$.  
Consider an eigenstate $\ket{\phi}$ of $\Pa\T$ corresponding to the eigenvalue $\gamma_{0}$. 
Using only the properties of the operators $\Pa\T$ and $\T$ one can readily show that
\begin{equation}
\ket{\phi} = (\Pa\T)^{2}\ket{\phi} = (\Pa\T)\gamma_{0}\ket{\phi} = \gamma_{0}^{*}\gamma_{0}\ket{\phi}.
\end{equation}
This means that $\gamma_{0}$ is a phase, and can be written as $\gamma_{0} = {e}^{i\alpha}$, with $\alpha \in \mathbb{R}$.
If we consider that $\ket{\phi}$ is also an eigenstate of the Hamiltonian $H$ with energy $E$ it is easy to show from the fact that $H$ and $\Pa\T$ commute that
\begin{equation}
E\gamma_{0}\ket{\phi} = E^{*}\gamma_{0}\ket{\phi}. 
\end{equation}
As $\gamma_{0}$ is nonvanishing it follows that $E$ is real.  
This conclusion breaks down when the eigenstates of $\Pa\T$ are \emph{not} the same as the eigenstates of the non-Hermitian Hamiltonian $H$. 
In this case we say that the $\Pa\T$-symmetry of $H$ is \emph{broken}, and conversely we refer to the case where the spectrum is real as the \emph{unbroken} case.  
It is important to note that the exact form of the operator $\Pa\T$ is not important in our discussion. 
When examining the spectrum of a non-Hermitian Hamiltonian, provided there is some antilinear operator $\Pa\T$ such that $(\Pa\T)^{2}=1$ and Eq.~\eqref{eq:symm} holds, then the spectrum is real or complex 
depending on whether or not the symmetry is broken.

%------------ Results-----------------------------%
\section{Warmup example: Single Spin system}
\label{sec:single}

\begin{figure*}
\includegraphics[width=2\columnwidth]{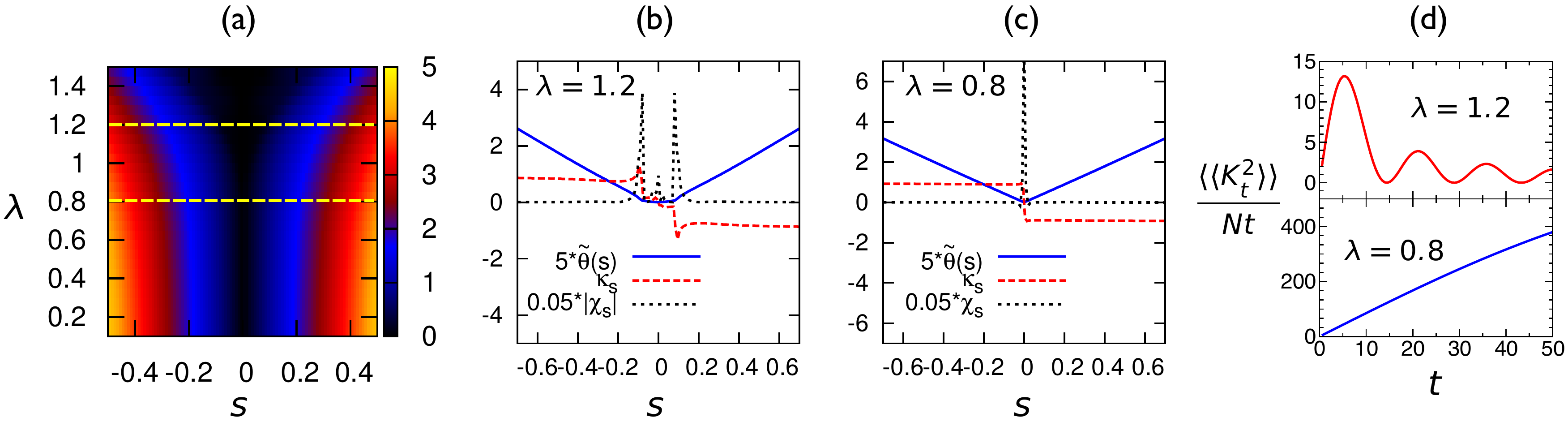}
\caption{(Color online) (a) Scaled CGF $\tilde{\theta}(s)$ in the $\lambda$-$s$ plane (from exact diagonalisation for chains of $N = 9$ spins).  For $\lambda > 1$ there is a large region where $\tilde{\theta}(s) \approx 0$.
(b,c) Plots of the scaled CGF, dynamical order parameter and susceptibility as a
function of $s$ for $\lambda = 0.8$ and $1.2$ ($N=11$ here).  We see that in the ferromagnetic
regime a large peak in the susceptibility at $s = 0$ that indicates the
cumulants grow faster than $t$.  In the paramagnetic regime the order parameter
is vanishing near $s = 0$, but beyond some critical $s$ value the $\Pa\T$-symmetry
breaks and it acquires a finite value.  This symmetry breaking is marked by peaks
in the susceptibility $\chi_{s}$.  (d) Scaled second cumulant of the time-integrated magnetisation in both regimes.}
\label{fig:fig2}
\end{figure*}

In this section we discuss a simple single spin Hamiltonian as a toy example of our approach of relating $\Pa\T$-symmetry to cumulants of time-integrated observables. 
Consider a single-spin system with Hamiltonian
\begin{equation}
\label{eq:singleH}
H = \epsilon \, \sigma^x ,
\end{equation}
where $\sigma^{x,y,z}$ denote the usual Pauli spin operators.  The operator of interest is the time-integrated magnetisation in a given direction in the $xz$ plane, i.e., as in Eq.\ \eqref{Kt} with 
\begin{equation}
k = \sigma^x \cos\varphi + \sigma^z \sin \varphi .
\end{equation}
Following the prescription of Sec.\ \ref{sec:Form}, the associated operator \eqref{eq:deform} becomes
\begin{equation}
\label{eq:singleHs}
H_s = \epsilon \, \sigma^x - \frac{is}{2}(\sigma^x \cos \varphi + \sigma^z \sin \varphi).
\end{equation}
This non-Hermitian Hamiltonian, discussed in a different context in Ref.~\cite{Bender2007}, has eigenvalues
\[
E_{\pm} = \pm\frac{1}{2}\sqrt{4\epsilon^{2}-s^{2}-4is\epsilon\cos\varphi} .
\] 
The $\Pa\T$ operator in this case is defined as the composition of $\Pa = \sigma^x$, and of $\T$, the operator indicating complex conjugation.

Consider first the case of the transverse magnetisation in the $z$ direction, i.e.,  $\varphi = \pi/2$. 
In this case $H_{s}$ is invariant under the joint action of $\Pa$ and $\T$.  The eigenvalues simplify to 
\begin{equation}
\label{eq:simpeig}
E_{\pm} = \pm\frac{1}{2}\sqrt{4\epsilon^{2}-s^{2}}.
\end{equation}
Provided $4\epsilon^{2}\ge s^{2}$, the eigenvalues are real and we are in the $\Pa\T$ {\em unbroken} regime.  
In this case one can use Eq.~\eqref{eq:MGF} to show that the cumulants of the time-integrated magnetisation in the $z$ direction oscillate in time, and are therefore sublinear at long times, as indicated by the fact that $\theta(s) = 0$.  When the $\Pa\T$-symmetry is \emph{broken}~\cite{Bender2007}, that is when $4\epsilon^{2} < s^{2}$, the eigenvalues form a complex conjugate pair, and the MGF takes the form
\begin{equation}
\label{eq:MGF2}
Z_{t}(s) = c_{+}(s) {e}^{2\text{I}\text{m}(E_+) t} + c_{-}(s)e^{2\text{I}\text{m}(E_-) t}.
\end{equation}
The coefficients $c_{\pm}$ are determined by the overlaps of the initial state with the eigenstates of $H_s$.  
In the long-time limit the MGF is dominated by the imaginary part of either $E_+$ or $E_-$, the choice of which depends on the sign of $s$. 
In fact a change of the sign of $s$ is equivalent to performing Hermitian conjugation on $H_s$.  
From Eq.~\eqref{eq:MGF2} one can easily see in this regime the CGF will scale linearly with $t$ as $t\rightarrow \infty$, and $\theta(s)$ will be 
finite.  This behaviour is illustrated in Fig.\ \ref{fig:fig1}(a) which shows both $\theta(s)$
and $\kappa_s$ as a function of $s$ in the case of $\varphi=\pi/2$: by tuning the counting field $s$ one may break the $\Pa\T$-symmetry, leading to square root singularities in $\theta(s)$ at $s = \pm 2\epsilon$.

If we consider instead the time-integrated magnetisation in any direction other than $z$, that is, for $\varphi \neq \pi/2$, then $H_{s}$ is not $\Pa\T$-symmetric for any parameter values.  In this case we are always in the 
regime where $\Pa\T$-symmetry plays no role, and the eigenvalues $E_{\pm}$ form a complex conjugate pair.   Note that this holds for all values of $s$.  
Considering $\Theta_{t}(s)/t$ as a function of $t$ we observe the emergence of a sharp feature at $s = 0$ that becomes more pronounced with increasing time.  This is due to the MGF being of the form given in Eq.~\eqref{eq:MGF2}, and in the long-time limit the eigenvalue which dominates is determined by 
the sign of $s$.  
This singularity implies that the cumulants grow faster than linear in $t$.  This is despite the fact that $\theta(s)$ is non-zero and well defined: the reason is that in general the limits 
$t \rightarrow \infty$ and $s \rightarrow 0$ do not necessarily commute.  
The emergence of this singularity in this parameter regime is shown in Fig.~\ref{fig:fig1}(c).  

This simple example of the single spin system illustrates that depending on the time-integrated observable $K_{t}$ being counted, and on the value of the counting field $s$, we can have regimes of unbroken and broken $\Pa\T$ symmetry which in turn determine the long-time behaviour of the cumulants $\llangle K_{t}^{n} \rrangle$.  In the next Section we study a many-body problem, the TFIM, and consider cumulants of the time-integrated longitudinal
magnetisation.

\section{Time-integrated longitudinal magnetisation in the Transverse field Ising model}
\label{sec:Ising}

In this section we consider the one-dimensional TFIM. 
This model is a paradigmatic example of a system that exhibits a static quantum phase transition~\cite{Sachdev2011} and is described by the Hamiltonian
\begin{equation}
H = - \sum_{i} \sigma_{i}^{x}\sigma_{i+1}^{x}-\lambda\sum_{i}\sigma_{i}^{z},
\end{equation}
where $i=1,\ldots,N$ indicate the sites of a one dimensional lattice with periodic boundary conditions.  
The Hamiltonian consists of two parts, the first is a bond operator term which seeks to align the spins (we have set the corresponding exchange coupling to one).  
The second term is the transverse field which acts against the bond operator. 
This Hamiltonian may be diagonalized using standard free fermion techniques~\cite{Sachdev2011}.
The system has a second order static quantum phase transition at $\lambda_{\rm c}= \pm 1$, where the ground state changes in a singular fashion from a ferromagnetic state at $-1<\lambda<1$ to a paramagnetic state elsewhere.

We wish to examine the cumulants of the time integrated total longitudinal magnetisation $M_{t}^{x}$. That is, the time-integrated operator $K_{t}$ we consider is
\begin{equation}
K_{t} = M_{t}^{x} \equiv \int_0^{t} \sum_{i} \sigma_{i}^{x}(t) ~ dt' , 
\end{equation}
where $\sigma_{i}^{x}(t)  = {e}^{itH} \sigma_{i}^{x} {e}^{-itH}$.  As explained above, in order to obtain the cumulants of $K_{t}$ we need to consider the deformed Hamiltonian, Eq.~\eqref{eq:deform}, from which we can extract the MGF,
\begin{equation}
\label{eq:HTFIM}
H_{s} = - \sum_{i}
\sigma_{i}^{x}\sigma_{i+1}^{x}-\lambda\sum_{i}\sigma_{i}^{z}-\frac{is}{2}\sum_{i}\sigma_{i}^{x}.
\end{equation}
This Hamiltonian corresponds to a discretization of the massive Yang-Lee model~\cite{Lee1952I,Lee1952II,Cardy1985,Belavin1984}.
Its critical properties have been studied extensively in~\cite{Gehlen1991}, where a critical line in the $s$-$\lambda$ plane was found.
This curve separates two regions in the parameter space where $H_s$ possesses a real spectrum and complex spectrum.
More recently \cite{Fring2009} the existence of these two regimes was explained analytically by analysing the appropriate $\Pa\T$-symmetry of $H_s$.

Following Ref.~\cite{Fring2009}, it is easy to show that the non-Hermitian Hamiltonian is $\Pa\T$-symmetric.  Specifically, if the $\Pa\T$ transformation is such that $\sigma_{i}^{x} \to - \sigma_{i}^{x}$ and $i \to -i$, then 
$H_{s} = H_{s}^{\Pa\T}$.  Furthermore, that there could be regimes of both broken and unbroken $\Pa\T$ can be seen form the following argument.  Under a rotation of $\pi/2$ around the $z$-axis, $\mathcal{R} \equiv {e}^{\frac{i\pi}{4}\sum_{i}\sigma_{i}^{z}}$, the transformed operator, $\mathcal{R}H_{s}\mathcal{R}^{-1}$, is real but non-symmetric.  This implies that its eigenvalues, and hence those of $H_{s}$, must either be real or appear in complex conjugate pairs, and thus that $\Pa\T$-symmetry may or may not be broken spontaneously. 
One is able then to diagonalize simultaneously $H_{s}$ and $\Pa\T$, and distinguish a region in which they possess the same eigenvectors (unbroken $\Pa\T$ regime), from a region in which they do not (broken $\Pa\T$ regime).
 
The $\Pa\T$-symmetry properties of $H_{s}$ have a direct impact on the time-integrated longitudinal magnetisation.
Lets consider in particular the case where the system is in its ground state, which we denote by $\ket{0}$ (generalisations to other pure or mixed states is straightforward).  In terms of the eigenvalues and eigenvectors of $H_{s}$ the MGF \eqref{eq:MGF} reads
\begin{equation}
Z_{t}(s) = \sum_{a,b} {e}^{i (E_{a}^{*}-E_{b})t}  \langle 0 \ket{L_{a}} \bra{R_{a}}R_{b}\rangle \bra{L_{b}}0\rangle ,
\label{ZZ}
\end{equation}
where $E_{a}$ ($a=1,\ldots,2^{N}$) is an eigenvalue of $H_s$, with associated right, $\ket{R_{a}}$, and left, $\bra{L_{a}}$, eigenvectors.  Since $H_{s}$ is non-Hermitian $E_{a}$ may be complex, and the left and right eigenvectors may not be simply related by Hermitian conjugation, $\ket{R_{a}}^{\dagger} \neq \bra{L_{a}}$.  In terms of these eigenvectors the orthogonality and completeness relations are, respectively, $\bra{L_{a}} R_{b} \rangle = \delta_{a,b}$ and $\mathbb{I} = \sum_{a} \ket{L_{a}} \bra{R_{a}}$.  The coefficients in 
\eqref{ZZ} relate to the overlap of the eigenstates of $H_s$ with the state of the system we are considering, in this case the ground state $\ket{0}$, and to the fact that due to the difference between left and right eigenstates 
$\ket{R_{a}}$ and $\ket{R_{b}}$ are not in general orthogonal.

In general, at long times, an MGF like (\ref{ZZ}) will be dominated by the term $a,b$ for which the imaginary part of $E_{b}-E_{a}^{*}$ is largest, provided that the overlaps $\langle 0 \ket{L_{a,b}}$ and $\bra{R_{a}}R_{b}\rangle$ are non-zero.  For the case of the time-integrated magnetisation in the TFIM we know that $H_{s}$ \eqref{eq:HTFIM} is $\Pa\T$-symmetric, and the eigenvalues are either real ($\Pa\T$-unbroken regime) or all come in complex conjugate pairs ($\Pa\T$-broken regime).  This means that the long-time (scaled) CGF will be
\begin{equation}
\tilde{\theta}(s) = \left\{
\begin{array}{cl}
0 & \text{$\Pa\T$-unbroken} \\
\max_{a} 2 \, \text{Im}(\varepsilon_{a}) & \text{$\Pa\T$-broken} \\
\end{array}
\right. ,
\end{equation}
where $\varepsilon_{a}$ is the complex eigenvalue per unit spin, $\varepsilon_{a} \equiv E_{a}/N$.  This in turn has direct implications for the time-integrated magnetisation.  In particular, in the $\Pa\T$-unbroken case the MGF \eqref{ZZ} is purely oscillatory, and so are all the cumulants $\llangle (M_{t}^{x})^{n} \rrangle$.  

We now confirm the above scenario by numerically determining $\tilde{\theta}(s)$.  We diagonalise $H_{s}$ \eqref{eq:HTFIM} for rings of $N=9$ and $11$ spins.  From Eq.\ \eqref{ZZ} we can obtain the associated (scaled) long-time CGF.  Figure~\ref{fig:fig2}(a) shows $\tilde{\theta}(s)$ in the $(\lambda,s)$-plane.  Even for these relatively small system sizes we find that there is a clear change of behaviour depending on the value of the transverse field $\lambda$.  For $\lambda < 1$ (static ferromagnetic phase) the CGF is non-zero for all values of $s\neq 0$.  In contrast, for $\lambda > 1$ (static paramagnetic phase) there is a finite range of values around $s=0$ for which $\tilde{\theta}(s) = 0$.  If we take slices at values of $\lambda$ in these two phases, Figures~\ref{fig:fig2}(b,c), we see that for $\lambda > 1$ the change from $\tilde{\theta}(s) = 0$ to $\neq 0$ is accompanied by a first-order jump in the order parameter $\kappa_{s}$ (and a peak in the susceptibility $\chi_{s}$).  In contrast, for $\lambda < 1$, the behaviour is similar to that of Fig.\ \ref{fig:fig1}(c).  The change in the behaviour of $\tilde{\theta}(s)$ is manifested also in a change in the time dependence of the cumulants $\llangle (M_{t}^{x})^{n} \rrangle$.  In Fig.\ \ref{fig:fig2}(d) we show the second cumulant, scaled by time and system size, 
$\llangle K_t^{2}\rrangle/(Nt)$, in both regimes: for the case $\lambda > 1$ it oscillates, and in the long time limit the scaled cumulant  vanishes; in contrast, for $\lambda < 1$, it grows in a super-linear fashion with time, and scaled cumulant grows with time.  Higher order cumulants behave similarly in the two regimes. 

\begin{figure}[h!]
\includegraphics[width=0.7\columnwidth]{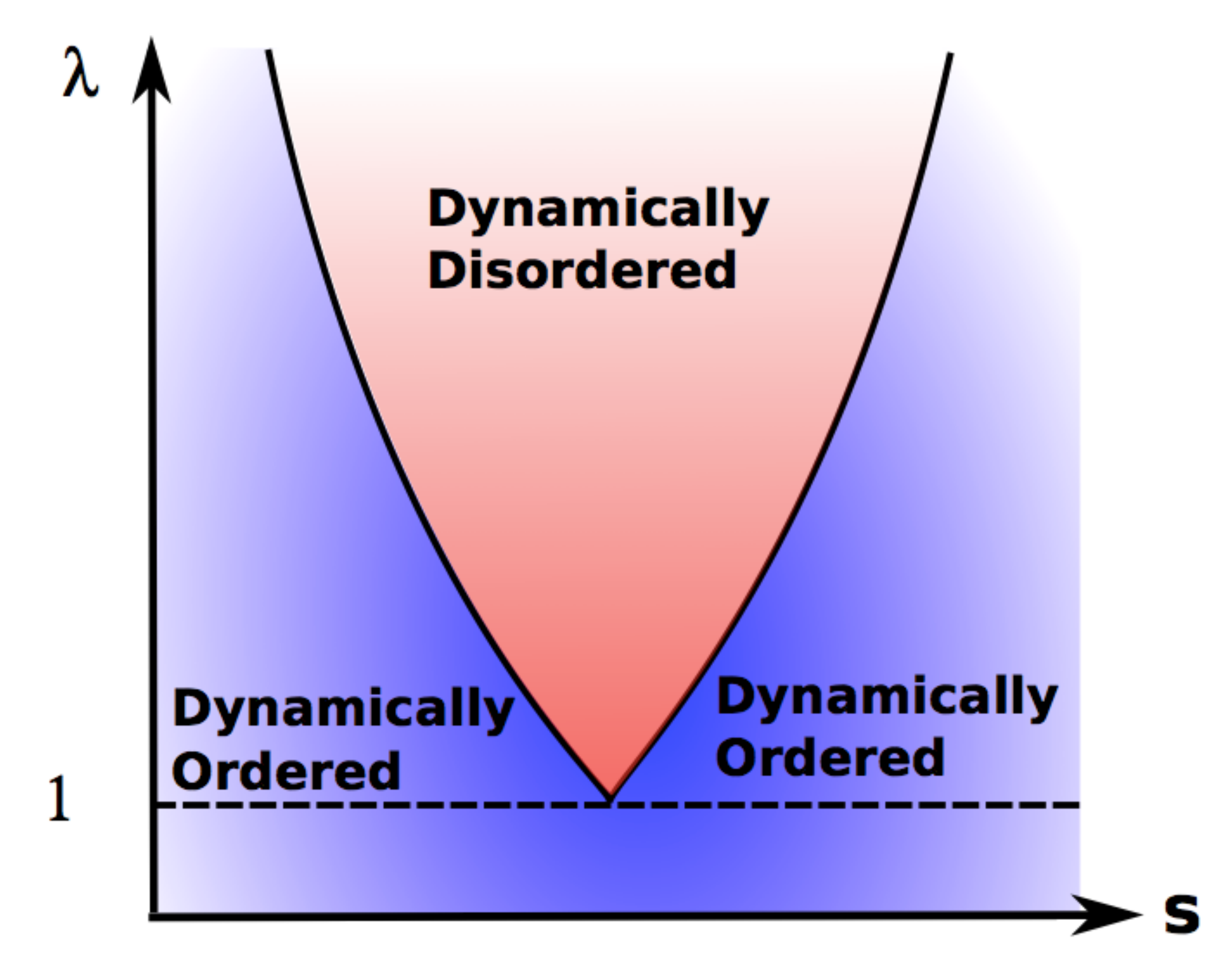}
\caption{(Color Online) Dynamical phase diagram of the TFIM, with time-integrated magnetisation $M_{t}^{x}$ as order parameter, in terms of the transverse field $\lambda$ and counting field $s$.  There are two dynamical phases: a ``dynamically ordered'' phase, which contains the static ordered phase of $|\lambda| < \lambda_{\rm c} = 1$, where the $\Pa\T$-symmetry of $H_{s}$ is broken, and as a consequence the cumulants of $M_{t}^{x}$ are linear or super-linear in time; and a ``dynamically disordered'' phase, which contains the static ordered phase of $|\lambda| > \lambda_{\rm c} = 1$, where $\Pa\T$-symmetry is unbroken, and as a consequence cumulants of $M_{t}^{x}$ are sub-linear in time.  The two phases are delimited by a curve of first-order transitions of $\theta(s)$ corresponding to the spontaneous breaking of $\Pa\T$-symmetry.
}
\label{fig:fig3}
 \end{figure}

In the thermodynamic limit, $N \to \infty$, the $\Pa\T$-symmetry breaking transition becomes sharp.  This defines a curve dynamical transitions in the $\lambda$-$s$ plane.   This dynamical phase-diagram is shown in Fig.\ \ref{fig:fig3}, and in fact coincides with that previously found in Ref.~\cite{Gehlen1991}.  The key difference now is that this curve, previously 
identified as critical, in the dynamical setup we consider corresponds to a curve of first-order transitions of the long-time CGF $\theta(s)$. That is, at these transition points the dynamical order parameter $\kappa_s$ changes discontinuously, cf.\ Fig.\ \ref{fig:fig2}(b).  
Note that these sharp changes in the dynamical features are not directly predicted just the statical properties.  For example, imagine preparing the system in the ground state at two values of $\lambda$ equidistant from the static critical point $\lambda_{\rm c}$.  These two states will have the same correlation length, which is determined by the distance from the quantum critical point, $|\lambda-\lambda_{\rm c}|$, but, as shown above, the behaviour of the cumulants of $M_t^x$ will be very different in these two states.

%------------ Conclusions --------------------%
\section{Discussion}
\label{sec:Conc}

We discussed the connection between the behaviour of a time-integrated observable in a closed quantum system and the possible $\Pa\T$-symmetry of the non-Hermitian Hamiltonian \cite{Hickey2013} that defines the generating functions for moments and cumulants of this observable.  We studied in particular the case of the transverse field Ising model in one dimension, and considered as a dynamical observable the time-integrated longitudinal magnetisation.  We showed that the associated generating functions are singular on a curve in the magnetic and counting field plane that separates regimes of distinct $\Pa\T$-symmetry properties.  In one regime 
$\Pa\T$-symmetry is spontaneously broken (a regime which includes the static ferromagnetic phase), and cumulants of the time-integrated magnetisation grow at least linearly in time.  In a second regime (which includes the static paramagnetic phase), $\Pa\T$-symmetry is unbroken, and cumulants are sublinear in time.  

We focussed here on the TFIM since its $\Pa\T$-symmetry properties when adding an imaginary perturbation, in particular an imaginary longitudinal field, are known \cite{Gehlen1991,Fring2009}.  But the dynamical implications of $\Pa\T$-symmetry could possibly be more general.  Consider a many-body system whose Hamiltonian is invariant under certain symmetry operations $\Sym$, that is, $[\Sym, H]=0$; lets say this is a discrete $Z_{2}$ symmetry as in the TFIM case for the sake of simplicity, but one can think of more general situations as well \cite{Sachdev2011}.  In general there will be a choice of system extensive operators $k$ which are odd under this symmetry, $\Sym^{-1} k \Sym = -k$.  If one then considers the time-integral of $k$ as a time-extensive observable, the associated non-Hermitian Hamiltonian \eqref{eq:deform} will be $\Pa\T$-symmetric, with $\Pa$ given by $\Sym$, and $\T$ by complex conjugation.  As discussed above, such non-Hermitian operators may display regimes of broken and unbroken $\Pa\T$-symmetry in their spectrum \cite{Gehlen1991,Fring2009}, with the consequent distinct behaviour of the cumulants of the time-integrated observable. This means that there is a likelihood that strongly interacting systems, in particular those where a symmetry $\Sym$ is directly related to a quantum phase transition in the ground state of $H$, will display the different dynamical phases associated with the long-time dynamics of cumulants as occurs in the TFIM.  We find particularly interesting the fact that in the $\Pa\T$-symmetric regime cumulants are sub-linear in time: in the case of open (i.e.\ dissipative or stochatic) systems, such behaviour is associated to an absence of decorrelation (since cumulants of time-integrated quantities become linear in time for times larger than the relaxation time of the time-correlations of their integrands).  It may be that the existence of such a dynamical regime has implications for the ``thermalisation'' of certain classes of operators in closed many-body systems \cite{Polkovnikov2011}.

\bigskip

\acknowledgements
This work was supported by EPSRC Grant no.~EP/I017828/1 and Leverhulme Trust
grant no.~F/00114/BG.

\end{document}